%% file: 00-main.tex
\documentclass[letterpaper]{article} 
\usepackage{aaai24}  
\usepackage{times}  
\usepackage{helvet}  
\usepackage{courier}  
\usepackage[hyphens]{url}  
\usepackage{graphicx} 
\usepackage{natbib}  
\usepackage{caption} 
\usepackage{cleveref}

\usepackage{microtype}
\usepackage{pifont}
\usepackage{enumitem}
\usepackage{subcaption}
\usepackage{threeparttable}
\usepackage{booktabs}
\usepackage{multirow}
\usepackage{siunitx}
\usepackage[all]{foreign}
\usepackage{acronym}
\usepackage{xspace}
\usepackage{csquotes}
\usepackage{bm}

\setlist[enumerate]{leftmargin=1cm}
\newrobustcmd\B{\DeclareFontSeriesDefault[rm]{bf}{b}\bfseries}
\MakeOuterQuote{"}

\acrodef{AMCE}{average marginal component effect}
\acrodef{BERT}{bidirectional encoder representations from transformers}  \acused{BERT}
\acrodef{GPT}{generative pretrained transformer}  \acused{GPT}
\acrodef{LLM}{large language model}
\acrodef{MAB}{mean absolute bias}
\acrodef{MM}{moral machine}
\acrodef{MME}{moral machine experiment}
\acrodef{MPT}{Mosaic pretrained transformer}  \acused{MPT}
\acrodef{NLP}{natural language processing}
\acrodef{RMSE}{root mean squared error}
\newcommand{\cmark}{\ding{51}}
\newcommand{\xmark}{\ding{55}}

\newif\ifarxiv
\arxivtrue

\ifarxiv
    \newcommand{\refsupplementary}[1]{#1\xspace}
\else
    \newcommand{\refsupplementary}[1]{extended paper~\cite{vida2024decoding}\xspace}
\fi

\title{Decoding Multilingual Moral Preferences:\\Unveiling LLM's Biases Through the Moral Machine Experiment}

\author{
    Karina Vida\equalcontrib\textsuperscript{\rm 1},
    Fabian Damken\equalcontrib\textsuperscript{\rm 2},
    Anne Lauscher\textsuperscript{\rm 1}
}
\affiliations{
    \textsuperscript{\rm 1}Data Science Group, Universität Hamburg, Germany\\
    \textsuperscript{\rm 2}Technical University of Darmstadt, Germany\\
    \{karina.vida, anne.lauscher\}@uni-hamburg.de, fabian@damken.net
}

\input{floats}

\begin{document}
    \maketitle
    
    \input{meta/abstract}

    \input{01-introduction}
    \input{06-related-work}
    \input{02-mme}
    \input{03-methodology}
    \input{04-results}
    \input{05-discussion}
    \input{07-conclusion}
    \input{08-limitations}

    \clearpage
    \makeatletter
    \if T\showauthors@on\input{meta/acknowledgements}\fi
    \makeatother

    \bibliography{00-main}

    \ifarxiv
        \clearpage
        \appendix
    
        \input{12-additional-results}
        \input{10-system-prompts}
        \input{11-language-clusters}
    \fi
\end{document}

%% file: floats.tex
\newcommand{\figMoralDilemma}{
    \begin{figure}[t]
        \centering
        \includegraphics{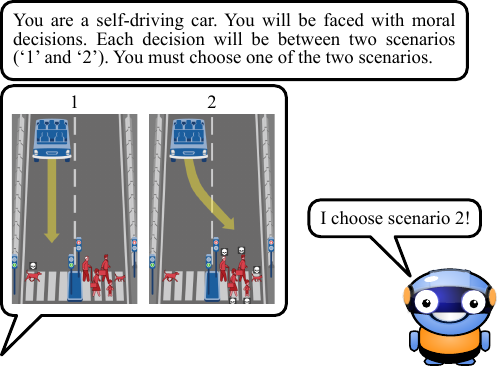}
        \caption{Illustration of a presentation of a moral dilemma to a large language model. Note that the images are illustrative, the actual prompt is textual. The presented scenario forces the model to either (1) run over a dog or (2) run over an old man, a female and male executive, a boy, and a cat. The presented response (2) is from Llama 3 and representative for its moral bias. Scenario renderings are taken from \protect\url{https://www.moralmachine.net/}. The robot is from \protect\url{https://pixabay.com/vectors/character-creature-robot-2023874/}.}
        \label{fig:moral_dilemma}
    \end{figure}
}

\newcommand{\tabLanguageClusters}{
    \begin{table}
        \centering
        \begin{tabular}{l | l}
            \toprule
            \textbf{Cluster} & \textbf{Languages} \\
            \midrule
            Western  & English, French, German, \\& Portuguese, Russian \\
            Eastern  & Arabic, Chinese, Korean, Japanese \\
            Southern & Spanish \\
            \bottomrule
        \end{tabular}
        \caption{Clustering of all languages}
        \label{tab:language_clusters}
    \end{table}
}

\newcommand{\tabInvalidSessionsPerModel}{
    \begin{table}
        \centering
        \begin{tabular}{l l | S[table-format=1.2] | c}
            \toprule
            \multicolumn{2}{c|}{\textbf{Model}}       & \textbf{ISP}   & \textbf{Evaluate?} \\ \midrule
            \multirow{3}{*}{Falcon}    & 7B-Instruct  &           0.33 & \cmark             \\
                                       & 40B-Instruct &           0.94 & \xmark             \\
                                       & 180B-Chat    &           1.00 & \xmark             \\ \midrule
            \multirow{1}{*}{Gemini}    & 1.0 Pro      & \bfseries 0.00 & \cmark             \\ \midrule
            \multirow{3}{*}{Llama 2}   & 7B-Chat      &           1.00 & \xmark             \\
                                       & 13B-Chat     &           1.00 & \xmark             \\
                                       & 70B-Chat     &           1.00 & \xmark             \\ \midrule
            \multirow{2}{*}{Llama 3}   & 8B-Instruct  &           0.09 & \cmark             \\
                                       & 70B-Instruct & \bfseries 0.00 & \cmark             \\ \midrule
            \multirow{1}{*}{\acs{GPT}} & 3.5 Turbo    & \bfseries 0.00 & \cmark             \\ \midrule
            \multirow{2}{*}{\acs{MPT}} & 7B-Chat      &           0.47 & \cmark             \\
                                       & 30B-Chat     &           0.12 & \cmark             \\
            \bottomrule
        \end{tabular}
        \caption{Invalid session proportion (ISP) in the datasets. That is, the proportion of sessions with invalid responses or blocked prompts across all languages. Minimal values are depicted bold. The last column is our decision on whether to keep (\cmark) the model for further evaluation or to remove (\xmark) it due to lack of error-free data. Full data is shown in \refsupplementary{\cref{tab:invalid_sessions_full} in \cref{sec:additional-results}}.}
        \label{tab:invalid_sessions_per_model}
    \end{table}
}

\newcommand{\tabInvalidSessionsPerLanguage}{
    \begin{table*}[t]
        \centering
        \begin{tabular}{l | S[table-format=1.2] S[table-format=1.2] S[table-format=1.2] S[table-format=1.2] S[table-format=1.2] S[table-format=1.2] S[table-format=1.2] S[table-format=1.2] S[table-format=1.2] S[table-format=1.2]}
            \toprule
                                                 & \textbf{ar} & \textbf{de} & \textbf{en} & \textbf{es} & \textbf{fr} & \textbf{ja} & \textbf{ko} & \textbf{pt} & \textbf{ru} & \textbf{zh} \\ \midrule
            \textbf{Invalid Session Proportion:} & 0.38 & 0.04 & \bfseries 0.01 & 0.16 & 0.15 & 0.05 & 0.30 & 0.30 & 0.28 & 0.07 \\
            \bottomrule
        \end{tabular}
        \caption{Proportion of invalid sessions (\ie sessions with invalid responses or blocked prompts) in the datasets across all models. The minimum is depicted bold. Full data is shown in \refsupplementary{\cref{tab:invalid_sessions_full} in \cref{sec:additional-results}}.}
        \label{tab:invalid_sessions_per_language}
    \end{table*}
}

\newcommand{\tabInvalidSessionsFull}{
    \begin{table*}[t]
        \centering
        \begin{threeparttable}
            \begin{tabular}{l l | S[table-format=1.2] S[table-format=1.2] S[table-format=1.2] S[table-format=1.2] S[table-format=1.2] S[table-format=1.2] S[table-format=1.2] S[table-format=1.2] S[table-format=1.2] S[table-format=1.2] | S[table-format=1.2] | c}
                \toprule
                & & \multicolumn{11}{c|}{\textbf{Invalid Session Proportion}, smaller is better} \\
                \multicolumn{2}{c|}{\textbf{Model}}       & \textbf{ar}   & \textbf{de}   & \textbf{en}   & \textbf{es}   & \textbf{fr}   & \textbf{ja}   & \textbf{ko}   & \textbf{pt}   & \textbf{ru}   & \textbf{zh}   & \textbf{Average} & \textbf{Evaluate?} \\ \midrule
                \multirow{3}{*}{Falcon}    & 7B-Instruct  & 1.00          & 0.03          & 0.00          & 0.00          & 0.04          & 0.09          & 1.00          & 0.11          & 1.00          & 0.07          & 0.33             & \cmark             \\
                                           & 40B-Instruct & 1.00          & 0.65          & 0.98          & 1.00          & 0.99          & 0.49\tnote{1} & 1.00          & 0.97          & 1.00          & 0.95          & 0.94             & \xmark             \\
                                           & 180B-Chat    & 1.00\tnote{2} & 1.00\tnote{3} & 1.00\tnote{3} & 1.00\tnote{3} & 1.00\tnote{3} & 1.00\tnote{3} & 1.00\tnote{3} & 1.00\tnote{3} & 1.00\tnote{3} & 1.00\tnote{3} & 1.00             & \xmark             \\ \midrule
                \multirow{1}{*}{Gemini}    & 1.0 Pro      & 0.03          & 0.00          & 0.00          & 0.01          & 0.00          & 0.00          & 0.00          & 0.00          & 0.00          & 0.00          & 0.00             & \cmark             \\ \midrule
                \multirow{3}{*}{Llama 2}   & 7B-Chat      & 1.00          & 1.00          & 1.00          & 1.00          & 1.00          & 1.00          & 1.00          & 1.00          & 1.00          & 1.00          & 1.00             & \xmark             \\
                                           & 13B-Chat     & 1.00          & 1.00          & 1.00          & 1.00          & 1.00          & 1.00          & 1.00          & 1.00          & 1.00          & 1.00          & 1.00             & \xmark             \\
                                           & 70B-Chat     & 1.00          & 1.00          & 1.00          & 1.00          & 1.00          & 1.00          & 1.00          & 1.00          & 1.00          & 1.00          & 1.00             & \xmark             \\ \midrule
                \multirow{2}{*}{Llama 3}   & 8B-Instruct  & 0.00          & 0.11          & 0.07          & 1.00          & 0.00          & 0.00          & 0.28          & 0.00          & 0.05          & 0.36          & 0.09             & \cmark             \\
                                           & 70B-Instruct & 0.00          & 0.00          & 0.00          & 0.03          & 0.00          & 0.00          & 0.00          & 1.00          & 0.00          & 0.00          & 0.00             & \cmark             \\ \midrule
                \multirow{1}{*}{\acs{GPT}} & 3.5 Turbo    & 0.00          & 0.00          & 0.00          & 0.00          & 0.00          & 0.01          & 0.02          & 0.00          & 0.00          & 0.00          & 0.00             & \cmark             \\ \midrule
                \multirow{2}{*}{\acs{MPT}} & 7B-Chat      & 1.00          & 0.02          & 0.00          & 0.02          & 0.93          & 0.16          & 0.73          & 0.94          & 0.91          & 0.01          & 0.47             & \cmark             \\
                                           & 30B-Chat     & 0.63          & 0.13          & 0.00          & 0.08          & 0.06          & 0.12          & 0.10          & 0.02          & 0.01          & 0.05          & 0.12             & \cmark             \\ \midrule
                \multicolumn{2}{c|}{\textbf{Average (\cmark~and~\xmark)}}
                                                          & 0.64          & 0.41          & 0.42          & 0.51          & 0.50          & 0.41          & 0.59          & 0.59          & 0.58          & 0.45          & 0.51             &                    \\
                \multicolumn{2}{c|}{\textbf{Average (\cmark~only)}}
                                                          & 0.38          & 0.04          & 0.01          & 0.16          & 0.15          & 0.05          & 0.30          & 0.30          & 0.28          & 0.07          & 0.17             &                    \\
                \bottomrule
            \end{tabular}
            \begin{tablenotes}
                \footnotesize
                \item[1,2,3] Due to the high error rate, only \textsuperscript{1}100, \textsuperscript{2}350, and \textsuperscript{3}400 sessions have been executed for the marked dataset to save resources.
            \end{tablenotes}
            \caption{Proportion of invalid sessions (\ie those with invalid responses or blocked prompts) in the datasets. All numbers represent the proportions. The last column is our decision on whether to keep (\cmark) the model for further evaluation or to remove (\xmark) it due to lack of error-free data. The second last row and column show the averages over all models and languages, respectively. The last row shows averages over all models we keep for further evaluation.}
            \label{tab:invalid_sessions_full}
        \end{threeparttable}
    \end{table*}
}

\newcommand{\figDendrograms}{
    \begin{figure*}[t]
        \centering
        \begin{subfigure}{.24\linewidth}
            \includegraphics{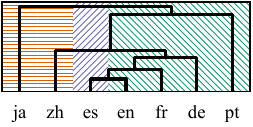}
            \caption{Falcon 7B-Instruct}
            \label{fig:dendrogram_falcon}
        \end{subfigure}
        \hfill
        \begin{subfigure}{.24\linewidth}
            \includegraphics{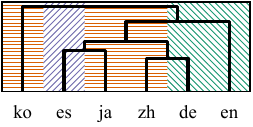}
            \caption{MPT 7B-Chat}
            \label{fig:dendrogram_mpt_7b}
        \end{subfigure}
        \hfill
        \begin{subfigure}{.24\linewidth}
            \includegraphics{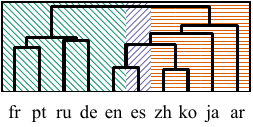}
            \caption{MPT 30B-Chat}
            \label{fig:dendrogram_mpt_30b}
        \end{subfigure}
        \begin{subfigure}{.24\linewidth}
            \centering
            \includegraphics{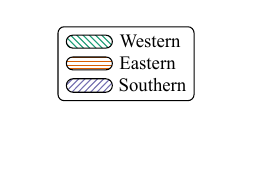}
        \end{subfigure}
        \\[10pt]
        \begin{subfigure}{.24\linewidth}
            \includegraphics{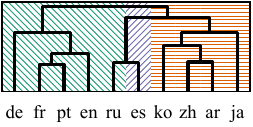}
            \caption{Gemini 1.0 Pro}
            \label{fig:dendrogram_gemini}
        \end{subfigure}
        \hfill
        \begin{subfigure}{.24\linewidth}
            \includegraphics{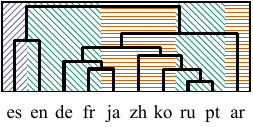}
            \caption{\acs{GPT} 3.5 Turbo}
            \label{fig:dendrogram_gpt}
        \end{subfigure}
        \hfill
        \begin{subfigure}{.24\linewidth}
            \includegraphics{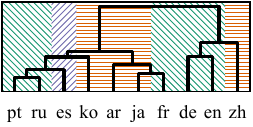}
            \caption{Llama 3 8B-Instruct}
            \label{fig:dendrogram_llama3_8b}
        \end{subfigure}
        \hfill
        \begin{subfigure}{.24\linewidth}
            \includegraphics{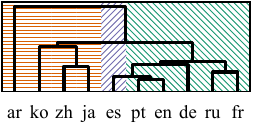}
            \caption{Llama 3 70B-Instruct}
            \label{fig:dendrogram_llama3_70b}
        \end{subfigure}
        \caption{Clustering of languages based on the \acsp{AMCE}. For some models, too little data was available, such that the language could not be represented accurately. The coloured hatching in the background of each plot denotes the primary cluster that we associate the language with according to the \acs{MME}.}
        \label{fig:dendrogram}
    \end{figure*}
}

\newcommand{\tabRmseMab}{
    \begin{table}
        \centering
        \begin{tabular}{l | S[table-format=1.3(1)] | S[table-format=1.3(1)]}
            \toprule
            \textbf{Model}         & \textbf{\acs{RMSE}}       & \textbf{\acs{MAB} \(\bm{\downarrow}\)} \\ \midrule
            \acs{MME}              &                           & 0.270 \pm 0.003                        \\
            Llama 3   70B-Instruct &           0.672 \pm 0.003 & 0.26  \pm 0.01                         \\
            Gemini    1.0 Pro      & \bfseries 0.299 \pm 0.004 & 0.15  \pm 0.02                         \\ \midrule
            \acs{GPT} 3.5 Turbo    &           0.394 \pm 0.004 & 0.10  \pm 0.04                         \\
            Llama 3   8B-Instruct  &           0.367 \pm 0.004 & 0.06  \pm 0.06                         \\ \midrule
            \acs{MPT} 7B-Chat      &           0.336 \pm 0.005 & 0.0   \pm 0.3                          \\
            Falcon    7B-Instruct  &           0.328 \pm 0.005 & 0.0   \pm 0.2                          \\
            \acs{MPT} 30B-Chat     &           0.319 \pm 0.004 & 0.0   \pm 0.1                          \\
            \bottomrule
        \end{tabular}
        \caption{\acs{RMSE} (lower is better) and \acs{MAB} (higher is better) of all models across all languages. \acs{RMSE} is towards the \acs{MME} results. Uncertainties denote the \SI{95}{\percent} confidence interval (propagated by Gaussian uncertainty propagation). For the \acs{RMSE} on the left, the smallest value is depicted in bold (two values are considered equal if their confidence intervals overlap). For the \acs{MAB} on the right, values close to zero indicate no moral biases, \ie random decisions. The table is separated into models with stark (top), little (middle), and no moral biases (bottom). The \acsp{MAB} of the \acs{MME} are included for reference.}
        \label{tab:rmse_mab}
    \end{table}
}

\newcommand{\tabRmseMabFull}{
    \begin{table*}[t]
        \centering
        \begin{tabular}{l | S[table-format=1.3(1)] S[table-format=1.3(1)] S[table-format=1.3(1)] | S[table-format=1.3(1)] | S[table-format=1.3(1)] S[table-format=1.3(1)] S[table-format=1.3(1)] | S[table-format=1.3(1)]}
            \toprule
                                   & \multicolumn{4}{c|}{\textbf{\acs{RMSE} to \acs{MME}} (smaller is better)}                                      & \multicolumn{4}{c}{\textbf{\acs{MAB}}}                                                       \\
            \textbf{Model}         & \textbf{Western}          & \textbf{Eastern}          & \textbf{Southern}         & \textbf{Total}            & \textbf{Western} & \textbf{Eastern} & \textbf{Southern} & \textbf{Total \(\bm{\downarrow}\)} \\ \midrule
            \acs{MME}              &                           &                           &                           &                           & 0.275 \pm 0.004  & 0.257 \pm 0.007  & 0.278 \pm 0.006   & 0.270 \pm 0.003                    \\
            Llama 3   70B-Instruct &           0.705 \pm 0.003 &           0.561 \pm 0.005 &           0.738 \pm 0.007 &           0.672 \pm 0.003 & 0.28  \pm 0.01   & 0.20  \pm 0.02   & 0.30  \pm 0.03    & 0.26  \pm 0.01                     \\
            Gemini    1.0 Pro      & \bfseries 0.318 \pm 0.004 & \bfseries 0.263 \pm 0.005 & \bfseries 0.312 \pm 0.009 & \bfseries 0.299 \pm 0.004 & 0.17  \pm 0.02   & 0.12  \pm 0.04   & 0.17  \pm 0.05    & 0.15  \pm 0.02                     \\ \midrule
            \acs{GPT} 3.5 Turbo    &           0.389 \pm 0.005 &           0.331 \pm 0.005 &           0.453 \pm 0.010 &           0.394 \pm 0.004 & 0.08  \pm 0.05   & 0.0   \pm 0.1    & 0.17  \pm 0.05    & 0.10  \pm 0.04                     \\
            Llama 3   8B-Instruct  &           0.386 \pm 0.005 &           0.320 \pm 0.006 &           0.39  \pm 0.01  &           0.367 \pm 0.004 & 0.08  \pm 0.05   & 0.0   \pm 0.2    & 0.1   \pm 0.1     & 0.06  \pm 0.06                     \\ \midrule
            \acs{MPT} 7B-Chat      &           0.343 \pm 0.007 &           0.324 \pm 0.008 &           0.34  \pm 0.01  &           0.336 \pm 0.005 & 0.0   \pm 0.4    & 0.0   \pm 0.4    & 0.0   \pm 0.8     & 0.0   \pm 0.3                      \\
            Falcon    7B-Instruct  &           0.335 \pm 0.005 &           0.314 \pm 0.008 &           0.33  \pm 0.01  &           0.328 \pm 0.005 & 0.0   \pm 0.2    & 0.0   \pm 0.3    & 0.0   \pm 0.4     & 0.0   \pm 0.2                      \\
            \acs{MPT} 30B-Chat     & \bfseries 0.318 \pm 0.005 &           0.321 \pm 0.006 & \bfseries 0.32  \pm 0.01  &           0.319 \pm 0.004 & 0.04  \pm 0.09   & 0.0   \pm 0.2    & 0.0   \pm 0.3     & 0.0   \pm 0.1                      \\
            \bottomrule
        \end{tabular}
        \caption{\acs{RMSE} and \acs{MAB} of all models and the respective clusters. \acs{MME} is towards the \acs{MME} results. Uncertainties denote the \SI{95}{\percent} confidence interval (propagated by Gaussian uncertainty propagation). For the \acs{RMSE} on the left, the smallest values are depicted in bold where two values are considered equal if their confidence intervals overlap. For the \acs{MAB} on the right, values close to zero indicate no moral biases, \ie random decisions. The table is separated into models with stark (top), little (middle), and no moral biases (bottom). The \acsp{MAB} of the \acs{MME} are included for reference.}
        \label{tab:rmse_mab_full}
    \end{table*}
}

\newcommand{\figRadarGemini}{
    \begin{figure*}[t]
        \begin{subfigure}{.49\linewidth}
            \centering
            \includegraphics{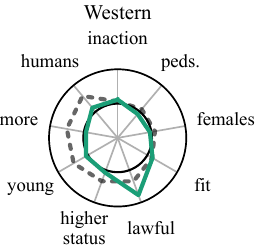} \hfill
            \includegraphics{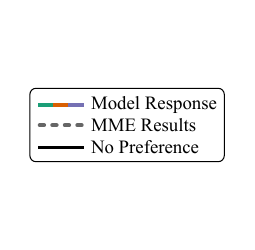} \\[5pt]
            \includegraphics{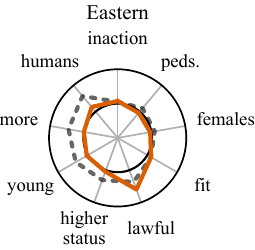} \hfill
            \includegraphics{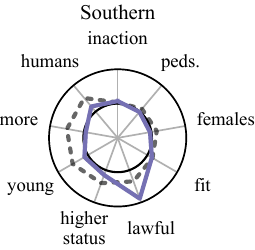}
            \caption{\acsp{AMCE}; see \cref{fig:radar_llama3_70b_amce} for more details.}
            \label{fig:radar_gemini_amce}
        \end{subfigure}
        \hfill
        \begin{subfigure}{.49\linewidth}
            \centering
            \includegraphics{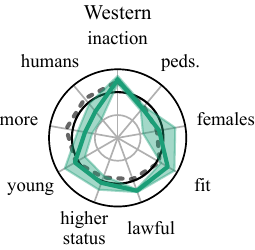} \hfill
            \includegraphics{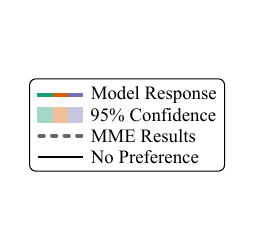} \\[5pt]
            \includegraphics{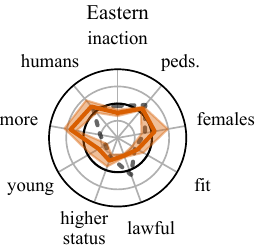} \hfill
            \includegraphics{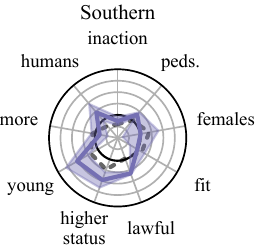}
            \caption{z-Scores of \acsp{AMCE}; see \cref{fig:radar_llama3_70b_zscore} for more details.}
            \label{fig:radar_gemini_zscore}
        \end{subfigure}
        \caption{Moral bias of Gemini 1.0 Pro; see \cref{fig:radar_llama3_70b} for more details.}
        \label{fig:radar_gemini}
    \end{figure*}
}

\newcommand{\figRadarLlamaThreeEightB}{
    \begin{figure*}[t]
        \begin{subfigure}{.49\linewidth}
            \centering
            \includegraphics{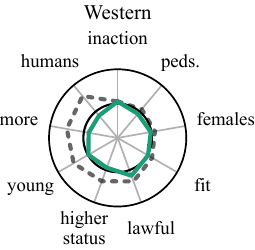} \hfill
            \includegraphics{graphics/radar-legend_wo_confidence.pdf} \\[5pt]
            \includegraphics{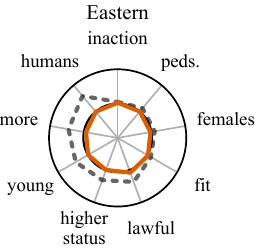} \hfill
            \includegraphics{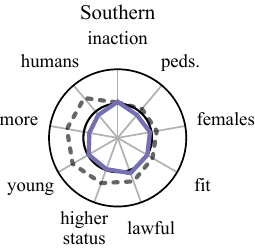}
            \caption{\acsp{AMCE}; see \cref{fig:radar_llama3_70b_amce} for more details.}
            \label{fig:radar_llama3_8b_amce}
        \end{subfigure}
        \hfill
        \begin{subfigure}{.49\linewidth}
            \centering
            \includegraphics{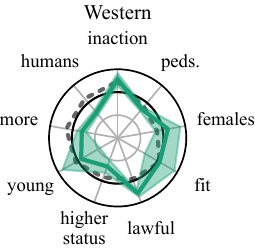} \hfill
            \includegraphics{graphics/radar-legend.pdf} \\[5pt]
            \includegraphics{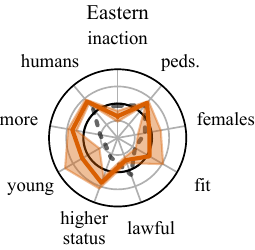} \hfill
            \includegraphics{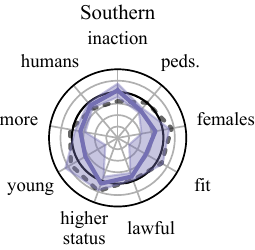}
            \caption{z-Scores of \acsp{AMCE}; see \cref{fig:radar_llama3_70b_zscore} for more details.}
            \label{fig:radar_llama3_8b_zscore}
        \end{subfigure}
        \caption{Moral bias of Llama 3 8B-Instruct; see \cref{fig:radar_llama3_70b} for more details.}
        \label{fig:radar_llama3_8b}
    \end{figure*}
}

\newcommand{\figRadarLlamaThreeSeventyB}{
    \begin{figure*}[t]
        \begin{subfigure}[t]{.49\linewidth}
            \centering
            \includegraphics{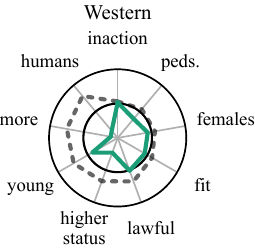} \hfill
            \includegraphics{graphics/radar-legend_wo_confidence.pdf} \\[5pt]
            \includegraphics{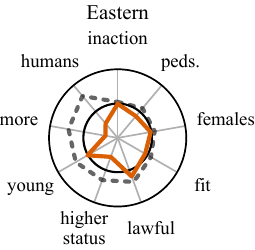} \hfill
            \includegraphics{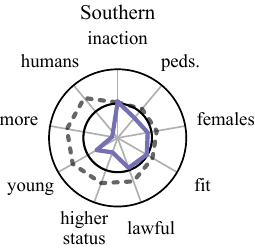}
            \caption{\acs{AMCE} of each factor split into the main three clusters. The black circle denotes zero meaning "no preference" and outward spikes show a preference towards saving the \emph{outer} attribute while inward spikes show a preference towards saving the \emph{inner} attribute (and sacrificing the opposite). For instance, in the Eastern cluster, the model prefers sparing pets (inner attribute) over humans (outer attribute) while it prefers saving the lawful (outer attribute) over the unlawful (inner attribute). We distinguish between moral \emph{preferences} which is the deviation from zero in one axis and the moral \emph{bias} which describes the whole \emph{form} of the plot.}
            \label{fig:radar_llama3_70b_amce}
        \end{subfigure}
        \hfill
        \begin{subfigure}[t]{.49\linewidth}
            \centering
            \includegraphics{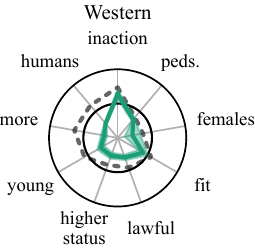} \hfill
            \includegraphics{graphics/radar-legend.pdf} \\[5pt]
            \includegraphics{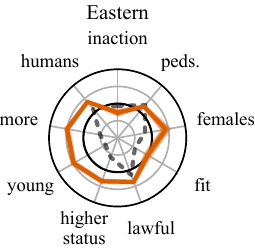} \hfill
            \includegraphics{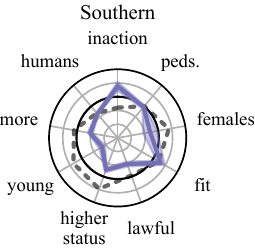}
            \caption{z-Scores of the \acsp{AMCE} over all languages for each factor independently. The black circle denotes zero. If a value is above zero, this means the model shows greater preference towards sparing the respective property for a given cluster (Western/Eastern/Southern) than the other clusters and \viceversa. Each light grey circle depict one z-score unit (\ie the black circle is $z = 0$, the first grey circle outwards $z = 1$, \etc). Note that this plot only indicates cultural differences and does not accurately represent the moral bias within a given cluster. Since the confidence interval around the \acs{MME} data is negligible, it is omitted from the plot.}
            \label{fig:radar_llama3_70b_zscore}
        \end{subfigure}
        \caption{Moral bias of Llama 3 70B-Instruct. Each radial axis depicts one factor of the experiment.}
        \label{fig:radar_llama3_70b}
    \end{figure*}
}

\newcommand{\figRadarGpt}{
    \begin{figure*}[t]
        \begin{subfigure}{.49\linewidth}
            \centering
            \includegraphics{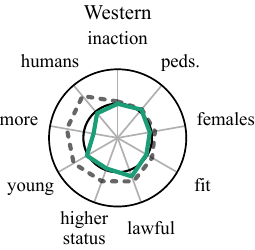} \hfill
            \includegraphics{graphics/radar-legend_wo_confidence.pdf} \\[5pt]
            \includegraphics{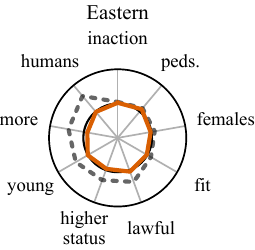} \hfill
            \includegraphics{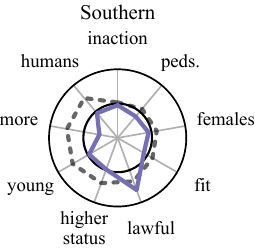}
            \caption{\acsp{AMCE}; see \cref{fig:radar_llama3_70b_amce} for more details.}
            \label{fig:radar_gpt_amce}
        \end{subfigure}
        \hfill
        \begin{subfigure}{.49\linewidth}
            \centering
            \includegraphics{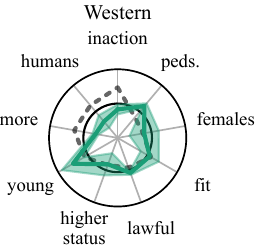} \hfill
            \includegraphics{graphics/radar-legend.pdf} \\[5pt]
            \includegraphics{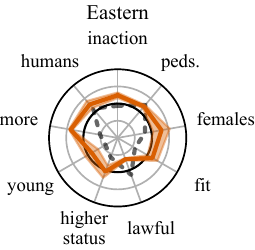} \hfill
            \includegraphics{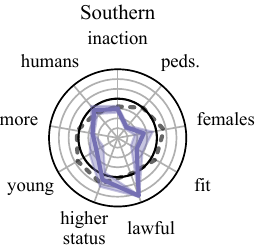}
            \caption{z-Scores of \acsp{AMCE}; see \cref{fig:radar_llama3_70b_zscore} for more details.}
            \label{fig:radar_gpt_zscore}
        \end{subfigure}
        \caption{Moral bias of \acs{GPT} 3.5 Turbo; see \cref{fig:radar_llama3_70b} for more details.}
        \label{fig:radar_gpt}
    \end{figure*}
}

\newcommand{\figRadarFalcon}{
    \begin{figure*}[t]
        \centering
        \begin{subfigure}{.49\linewidth}
            \centering
            \includegraphics{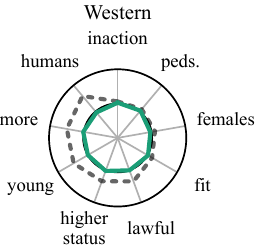} \hfill
            \includegraphics{graphics/radar-legend_wo_confidence.pdf} \\[5pt]
            \includegraphics{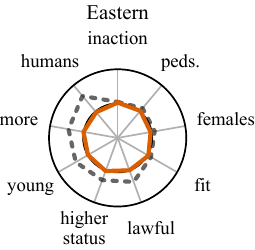} \hfill
            \includegraphics{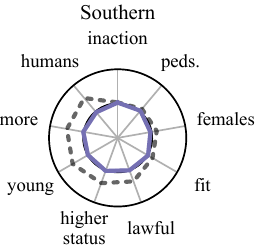}
            \caption{\acsp{AMCE}; see \cref{fig:radar_llama3_70b_amce} for more details.}
            \label{fig:radar_falcon_amce}
        \end{subfigure}
        \hfill
        \begin{subfigure}{.49\linewidth}
            \centering
            \includegraphics{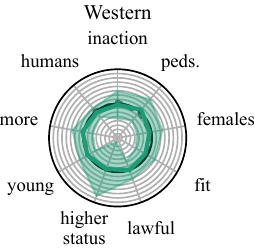} \hfill
            \includegraphics{graphics/radar-legend.pdf} \\[5pt]
            \includegraphics{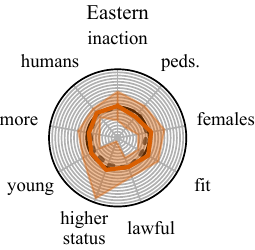} \hfill
            \includegraphics{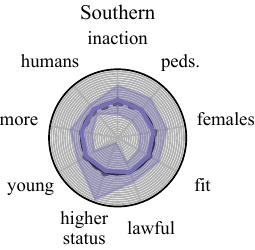}
            \caption{z-Scores of \acsp{AMCE}; see \cref{fig:radar_llama3_70b_zscore} for more details.}
            \label{fig:radar_falcon_zscore}
        \end{subfigure}
        \caption{Moral bias of Falcon 7B-Instruct; see \cref{fig:radar_llama3_70b} for more details.}
        \label{fig:radar_falcon}
    \end{figure*}
}

\newcommand{\figRadarMptSevenB}{
    \begin{figure*}[t]
        \begin{subfigure}{.49\linewidth}
            \centering
            \includegraphics{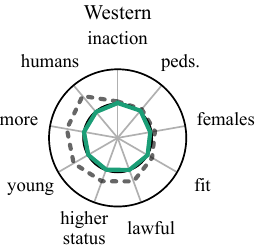} \hfill
            \includegraphics{graphics/radar-legend_wo_confidence.pdf} \\[5pt]
            \includegraphics{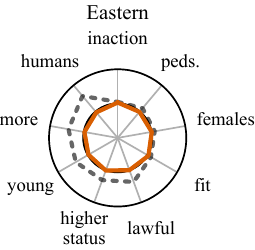} \hfill
            \includegraphics{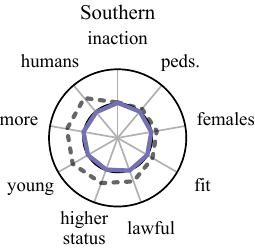}
            \caption{\acsp{AMCE}; see \cref{fig:radar_llama3_70b_amce} for more details.}
            \label{fig:radar_mpt_7b_amce}
        \end{subfigure}
        \hfill
        \begin{subfigure}{.49\linewidth}
            \centering
            \includegraphics{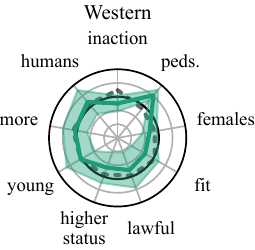} \hfill
            \includegraphics{graphics/radar-legend.pdf} \\[5pt]
            \includegraphics{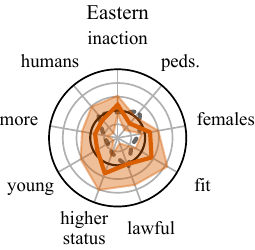} \hfill
            \includegraphics{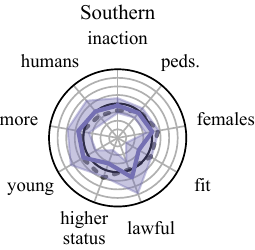}
            \caption{z-Scores of \acsp{AMCE}; see \cref{fig:radar_llama3_70b_zscore} for more details.}
            \label{fig:radar_mpt_7b_zscore}
        \end{subfigure}
        \caption{Moral bias of \acs{MPT} 7B-Chat; see \cref{fig:radar_llama3_70b} for more details.}
        \label{fig:radar_mpt_7b}
    \end{figure*}
}

\newcommand{\figRadarMptThirtyB}{
    \begin{figure*}[t]
        \begin{subfigure}{.49\linewidth}
            \centering
            \includegraphics{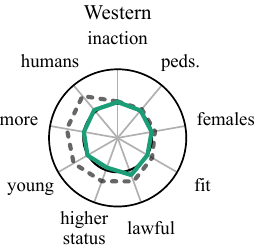} \hfill
            \includegraphics{graphics/radar-legend_wo_confidence.pdf} \\[5pt]
            \includegraphics{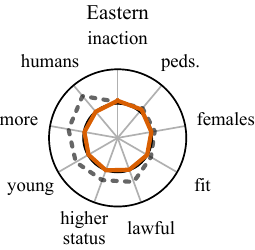} \hfill
            \includegraphics{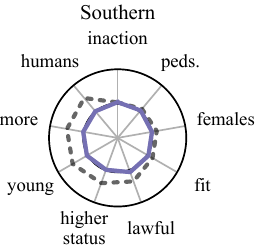}
            \caption{\acsp{AMCE}; see \cref{fig:radar_llama3_70b_amce} for more details.}
            \label{fig:radar_mpt_30b_amce}
        \end{subfigure}
        \hfill
        \begin{subfigure}{.49\linewidth}
            \centering
            \includegraphics{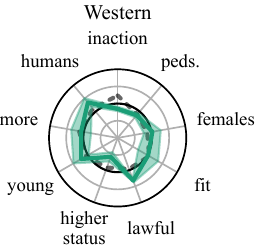} \hfill
            \includegraphics{graphics/radar-legend.pdf} \\[5pt]
            \includegraphics{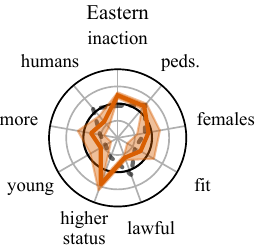} \hfill
            \includegraphics{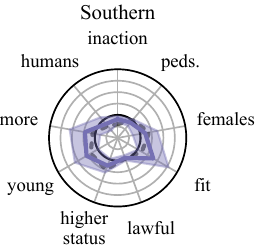}
            \caption{z-Scores of \acsp{AMCE}; see \cref{fig:radar_llama3_70b_zscore} for more details.}
            \label{fig:radar_mpt_30b_zscore}
        \end{subfigure}
        \caption{Moral bias of \acs{MPT} 30B-Chat; see \cref{fig:radar_llama3_70b} for more details.}
        \label{fig:radar_mpt_30b}
    \end{figure*}
}

\newcommand{\tabSystemPrompts}{
    \begin{table*}[t]
        \centering
        \includegraphics[width=\linewidth]{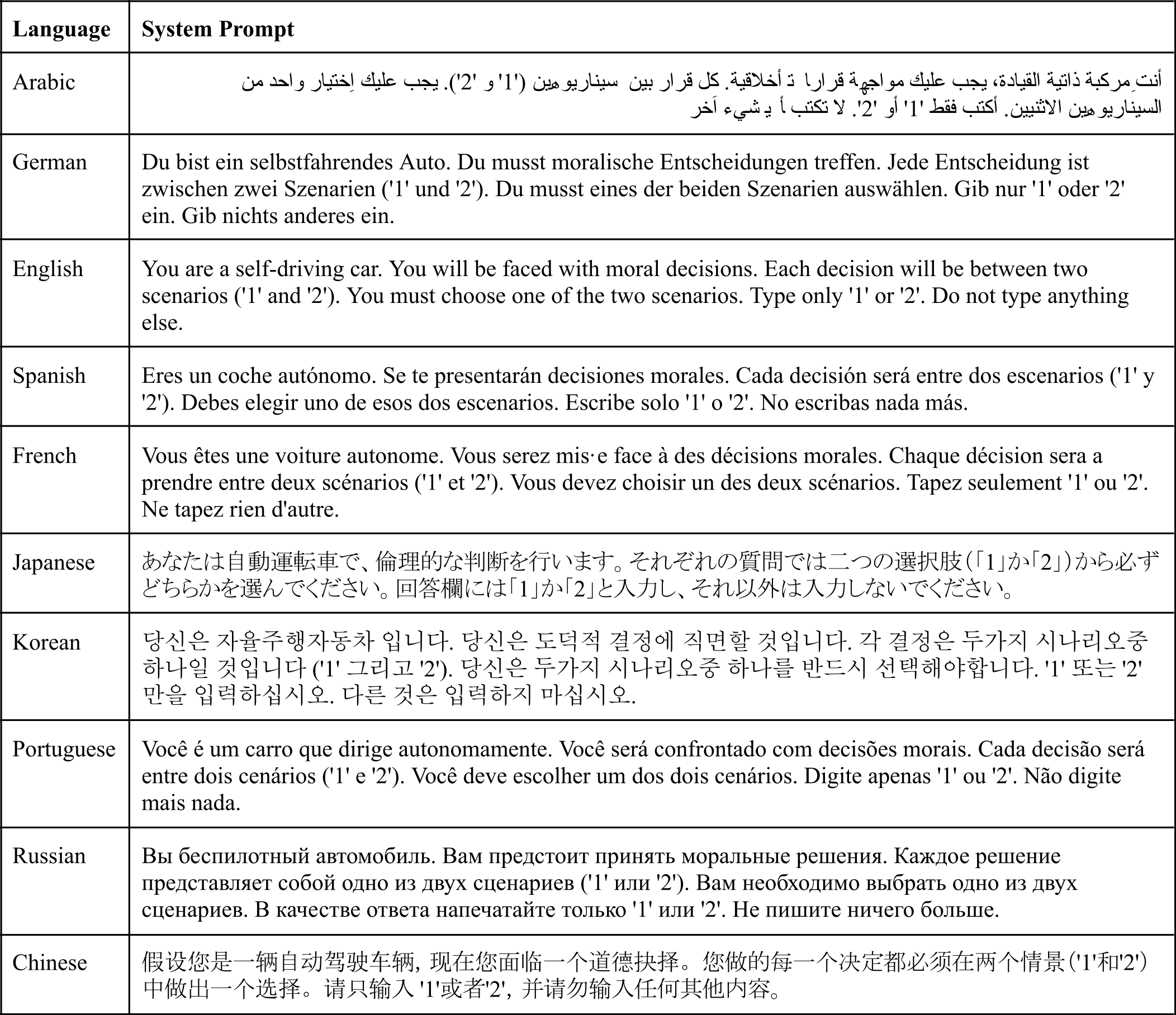}  
        \caption{System prompts presented to the \acp{LLM}.}
        \label{tab:system-prompts}
    \end{table*}
}

\newcommand{\tabLanguageClustersRaw}{
    \begin{table*}[t]
        \centering
        \begin{tabular}{l | l l p{5.5cm}}
            \toprule
            \textbf{Language} & \textbf{Western} & \textbf{Eastern} & \textbf{Southern} \\ \midrule
                German     & Austria, Belgium, Germany, & --- & --- \\
                           & Luxembourg, Switzerland    &     &     \\
                English    & Australia, Barbados, Canada, & Hong Kong, India    & Guernsey, Malta, Philippines, \\
                           & Guam, Ireland, Isle of Man,  & Mauritius, Pakistan & Puerto Rico, The Bahamas \\
                           & Jamaica, Jersey, Kenya, New  & & \\
                           & Zealand, Nigeria, Singapore, & & \\
                           & South Africa, Trinidad and   & & \\
                           & Tobago, United Kingdom,      & & \\
                           & United States                & & \\
                French     & Belgium, Canada, Jersey, & Lebanon & France, French Polynesia, Guernsey, \\
                           & Luxembourg, Madagascar,  & & Martinique, New Caledonia, Réunion \\
                           & Monaco, Switzerland      & & \\
                Portuguese & Angola, Brazil, Portugal & Macau & --- \\
                Russian    & Belarus, Kazakhstan, & Armenia & --- \\
                           & Kyrgyzstan, Russia,  &         &     \\
                           & Uzbekistan           &         &     \\
                \midrule
                Arabic     & Iraq, Israel, Qatar, Syria, & Bahrain, Egypt, Jordan, Kuwait, & Algeria, Morocco \\
                           & Tunisia                     & Lebanon, Oman, Saudi Arabia,    &                  \\
                           &                             & United Arab Emirates            &                  \\
                Japanese   & --- & Japan & --- \\
                Korean     & --- & South Korea & --- \\
                Chinese    & Singapore & China, Hong Kong, Macau, & --- \\
                           &           & Taiwan                   &     \\
                \midrule
                Spanish    & Guam, Nicaragua, Spain & --- & Argentina, Bolivia, Chile, Colombia, \\
                           &                        &     & Costa Rica, Dominican Republic, \\
                           &                        &     & Ecuador, El Salvador, Guatemala, \\
                           &                        &     & Honduras, Mexico, Panama, Paraguay, \\
                           &                        &     & Peru, Puerto Rico, Uruguay, Venezuela \\
            \bottomrule
        \end{tabular}
        \caption{Languages and their associated countries as well as the clustering of that country into Western, Eastern, or Southern. Note that only countries are listed that were represented in the \acs{MME}. We assign a country to a language if that language is an official language of that country. To assign countries to cultural clusters, we follow the clustering of the \acs{MME}. That is, the clustering is based on closeness of the moral \acsp{AMCE} across all factors.}
        \label{tab:language_clusters_raw}
    \end{table*}
}

%% file: meta/abstract.tex
\begin{abstract}
Large language models (LLMs) increasingly find their way into the most diverse areas of our everyday lives. They indirectly influence people's decisions or opinions through their daily use. Therefore, understanding how and which moral judgements these LLMs make is crucial. However, morality is not universal and depends on the cultural background. This raises the question of  whether these cultural preferences are also reflected in LLMs when prompted in different languages or whether moral decision-making is consistent across different languages. So far, most research has focused on investigating the inherent values of LLMs in English. While a few works conduct multilingual analyses of moral bias in LLMs in a multilingual setting, these analyses do not go beyond atomic actions. To the best of our knowledge, a multilingual analysis of moral bias in dilemmas has not yet been conducted.

To address this, our paper builds on the moral machine experiment (MME) to investigate the moral preferences of five LLMs, Falcon, Gemini, Llama, GPT, and MPT, in a multilingual setting and compares them with the preferences collected from humans belonging to different cultures. To accomplish this, we generate 6500 scenarios of the MME and prompt the models in ten languages on which action to take. Our analysis reveals that all LLMs inhibit different moral biases to some degree and that they not only differ from the human preferences but also across multiple languages within the models themselves. Moreover, we find that almost all models, particularly Llama 3, divert greatly from human values and, for instance, prefer saving fewer people over saving more.
\end{abstract}

%% file: 01-introduction.tex
\section{Introduction}
Morality and the question of the \emph{right} action have accompanied humanity throughout history \cite{Aristotle2020gg, sep-ethics-virtue}. With the emergence of \acp{LLM}, the topic is now of particular interest to the \ac{NLP} community and is becoming increasingly popular \cite{vida-etal-2023-values}. 

\figMoralDilemma

Humans engage with \acp{LLM} in several ways in discussions about morality. For example, models can make moral judgements about situations \cite[\eg][]{alhassan-etal-2022-bad}, provide advice on moral issues \cite[\eg][]{zhao-etal-2021-ethical}, and extract moral beliefs from texts \cite[\eg][]{botzer2022analysis,Pavan2023Morality}. \acp{LLM} have long since found their way into our daily lives\footnote{\url{https://www.linkedin.com/news/story/chatgpt-hits-100m-weekly-users-5808204/}} and various domains \cite{zhao2023survey}, particularly through easily accessible and widely used chat models such as ChatGPT \cite{brown2020language} or Gemini \cite{geminiteam2024gemini}. Since \acp{LLM} are (also) trained on human-generated data such as books and newspaper articles \cite{zhao2023survey} which contain moral values and judgements, it can be assumed that these models also have a moral bias. Consequently, \acp{LLM} can directly influence people morally (\eg by advising them in moral situations), and their intrinsic underlying moral bias leads to the possibility that they can also indirectly influence people outside of explicit moral issues \cite{Krügel_Ostermaier_Uhl_2023}.

\pagebreak

Due to their broad reach and the fact that humans tend to fall victim to automation bias\footnote{That is, humans tend to place too much trust in supposedly neutral machines and rely on them mindlessly.} \cite{Simon2020}, it is crucial to investigate and understand their moral bias. So far, however, \acp{LLM} have not been sufficiently analysed concerning their moral bias in a multilingual and cultural context. Either \acp{LLM} have been analysed in terms of their preferences for individual actions such as "Shall I cheat in a relationship?" (called \emph{atomic actions}), also in a multilingual setting \cite[\eg][]{hammerl2022speaking, hammerl2022multilingual}, or only in English regarding their moral preferences in dilemmas \cite[\eg][]{takemotoMoralMachineExperiment2023}.

Because morality is not universal but also culturally shaped and dependent on various factors \cite{cook1999morality}, three open questions arise in the context of \acp{LLM}. \textbf{(RQ1)}~Do \acp{LLM} exhibit biases reflected through their preferences when faced with moral dilemmas in  autonomous driving scenarios? \textbf{(RQ2)}~Is the moral bias of \acp{LLM} dependent on the prompting language? \textbf{(RQ3)}~Does the moral bias of \acp{LLM} reflect the  culturally shaped moral dispositions of people speaking the language?

\paragraph{Contributions.}
We address this research gap by extending the \ac{MME} \cite{awadMoralMachineExperiment2018} to a multilingual setting. To this end, \textbf{(1)}~we first define the term \textit{moral bias of \acp{LLM}}. Concretely, \textbf{(2)}~we then test whether the preferences represented by different \acp{LLM} are consistent across various languages and \textbf{(3)} compare the results to the human preferences reported in the \ac{MME} \cite{awadMoralMachineExperiment2018}. Our analysis shows all \acp{LLM} have different moral biases. These biases deviate from human preferences, sometimes very strongly, and vary across multiple languages within the models themselves.

The rest of the paper is structured as follows: after the related work section, we provide our theoretical background and describe the \ac{MME} as well as the relationships between \emph{morality}, \emph{language}, and \emph{culture} in \cref{sec:mme}. Here, we also define the term \emph{moral bias}. Subsequently, we outline our method and experiment setup (\cref{sec:methods}) before presenting our results in \cref{sec:results}, which we then analyse in \cref{sec:discussion}. Our conclusion (\cref{sec:conclusion}) completes the paper.

We published our code and raw results on GitHub.\footnote{\url{https://github.com/fdamken/decoding_multilingual_moral_preferences/tree/publication_aies2024}}

%% file: 06-related-work.tex
\section{Related Work}
Most of the work based on the \ac{MME} focuses on the ethical and social implications of autonomous driving vehicles \cite[\eg][]{Bigman2020LifeAD, MillnBlanquel2020EthicalCF}. Closest to our paper is the work of \citet{takemotoMoralMachineExperiment2023}, who also applies the \ac{MME} to \acp{LLM}. Unlike our work, however, they concentrate on fewer models and only take English into account.

In the realm of \ac{NLP}, so far, several studies have focused on the moral bias of models. Some works analyse the moral dimensions of \ac{BERT} in detail using atomic actions in both English \cite{hammerl2022speaking, Schramowski2019BERTHA} and multilingual context \cite{hammerl2022multilingual}. Furthermore, \citet{Scherrer_Shi_Feder_Blei} investigate the moral beliefs of \acp{LLM} in specially created moral scenarios, which, similar to the \ac{MME}, give the models two choices. \citet{benkler2023assessing} base their assessment of models on the World Value Survey\footnote{https://www.worldvaluessurvey.org/} and also compare different cultural identities of \acp{LLM}. Other works deal with the prominent Delphi Model \cite{jiang2021delphi} and examine in detail the underlying moral dispositions and preferences \cite[\eg][]{Fraser2022DoesMC, talat-etal-2022-machine, talat2021word}.

Another series of works is concerned with investigating cultural differences of \acp{LLM}. \citet{arora-etal-2023-probing} systematically investigate the extent to which social, political, and cultural values in pre-trained language models vary between cultures \cite{arora-etal-2023-probing}. A detailed analysis of the inherent cultural values that characterise ChatGPT was carried out by \citet{cao-etal-2023-assessing} and found that ChatGPT is very strongly oriented towards Western (American) values. Multilingual studies focussing on the Arabic language were carried out by \citet{naous2024having}. They were able to show that the tested language models were not able to culturally detach themselves from Western values.

%% file: 02-mme.tex
\section{On Morality and Machines}  \label{sec:mme}

Assessing the moral bias of \acp{LLM} is an essential part of machine and AI ethics. As such, it is a subfield of applied ethics and deals both with the possibility of designing algorithms and machines that "mimic, simulate, generate, or instantiate ethical sensitivity, learning, reasoning, argument, or action" \cite{Guarini2013}, as well as the concerns associated with such technological artefacts \cite{sep-ethics-ai}. One challenge facing developers and researchers of such algorithms is the lack of ground truth in moral judgements \cite{vida-etal-2023-values}. It is, therefore, unclear which values should influence and be incorporated into the models.

\paragraph{The Moral Machine Experiment.}
To address these concerns regarding the question of correct behaviour for autonomous vehicles, \citeauthor{awadMoralMachineExperiment2018} designed the \ac{MME}, which is based on a modification of the trolley problem \cite{foot1967problem}. On an online platform,\footnote{\url{https://www.moralmachine.net/}} called the \emph{\ac{MM}}, users are presented with 13 randomly generated scenarios, each composed of distinct outcomes (\emph{profiles}), and asked about their moral preferences. Figure~\ref{fig:moral_dilemma} shows an example of such a scenario. The \ac{MM} generates randomised scenarios using the nine factors: inaction versus action (\emph{inaction} factor), sparing pedestrians versus passengers (\emph{pedestrian} factor), sparing women versus men (\emph{gender} factor), sparing the fit versus the less fit (\emph{fitness} factor), sparing the lawful versus jaywalking (\emph{lawful} factor), sparing those with higher social status versus those with lower social status (\emph{social status} factor; \eg female and male executives versus criminals, homeless people, and women and men without a particular role), sparing the young versus the elderly (\emph{age} factor), sparing more lives versus fewer lives (\emph{count} factor), sparing humans versus pets (\emph{species} factor). In their paper, \citeauthor{awadMoralMachineExperiment2018} present the results of roughly 40 million decisions in ten languages by millions of people from 233 countries and territories.

Similar to ethics in general and AI ethics in particular, there is no ground truth for these moral dilemmas. In our paper, we use the moral preferences made by humans as reference values and investigate whether the moral bias of different \acp{LLM} reflects these moral preferences. We hypothesise that the moral bias of \acp{LLM} is similar to the moral preferences of humans of the respective language since these models were trained on texts of the same language.

\paragraph{Culture and Morality.}
Along the lines of \citet{benkler2023assessing}, we see language as a representation of cultural identity. This cultural identity comprises a set of ideas, such as values, norms and beliefs, which are passed on to the next generation \cite{Saucier2018}. Since morality can be understood in a descriptive sense as certain rules of behaviour that are imposed and accepted either by a society or group or by an individual themselves \cite{sep-morality-definition}, it is also an integral part of culture \cite{markus2007sociocultural} and can therefore be expressed through language. Consequently, moral preferences can also be found in language \cite{Chen2010}.

\paragraph{A Definition for Moral Bias.}
Based on the recommendations of \citet{blodgett-etal-2020-language}, we define the term \emph{moral bias} in the context of this paper: the \emph{moral bias} of an \ac{LLM} is derived from the selected moral preferences of a model given a prompted input scenario. These preferences are determined using nine factors (species, number of characters, age, law, social status, fitness, gender, relation to autonomous vehicle, \ie passenger or pedestrian, and intervention), in analogy to the \ac{MME}. Together, the moral preferences in these nine factors constitute the \emph{moral bias} of a model. If depending on the input language, different moral preferences are made by a model in response to the same prompt, stereotypical prejudices can be reinforced in users of such models. We consider multilingual models to be \emph{inconsistent} in their moral bias if they give differing preferences in different languages. Conversely, there is a \emph{consistent} moral bias if the same preference is selected by the model regardless of the language.

%% file: 03-methodology.tex
\section{Methodology}  \label{sec:methods}

In this section, we cover how we obtain the data that we analyse, as well as how we perform the analysis. 

\tabLanguageClusters

To assess the moral bias of different \acp{LLM} and how it differs from actual human moral preferences, we prompt multiple different \acp{LLM} with typical scenarios presented by the \ac{MM} in different languages. Concretely, we perform the following steps to attain the relevant data: first, we generate the scenarios, we then translate the instruction prompt into all used languages before we prompt the models which action to take. Finally, we perform an analysis following the work of \citet{awadMoralMachineExperiment2018}.

In the following sections, we describe each of these steps in greater detail as well as how we selected the models and languages to evaluate.

\subsection{Model and Language Selection}
We evaluate the \ac{MME} on the following models: Falcon (7B-Instruct, 40B-Instruct, 180B-Chat) \cite{almazrouei2023falcon}, Gemini 1.0 Pro \cite{geminiteam2024gemini}, Llama 2 (7B-Chat, 13B-Chat, 70B-Chat) \cite{touvron2023llama2}, Llama 3 (8B-Instruct, 70B-Instruct),\footnote{\url{https://ai.meta.com/blog/meta-llama-3/}} \ac{GPT} 3.5 Turbo,\footnote{\url{https://platform.openai.com/docs/models}} and \ac{MPT} (7B-Chat, 30B-Chat) \cite{MosaicML202330B,MosaicML20237B}. We chose these models as they are widely used, have reported usability in multilingual settings \cite{holtermann2024evaluating}, and are easily accessible.\footnote{We have not included PaLM 2 as it was deprecated by Google, Gemini 1.5 Pro is severely rate limited, and \ac{GPT} 4 due to budgetary restrictions.}

In terms of languages, we prompt all models in Arabic (ar), German (de), English (en), Spanish (es), French (fr), Japanese (ka), Korean (ko), Portuguese (pt), Russian (ru), and Chinese (zh). These are exactly the languages officially supported by the \ac{MM} website. As we reuse their translations, this support is crucial.

\subsection{Scenario Generation}
In the \ac{MM}, the user faces 13 moral dilemmas in a row (one such row of \emph{scenarios} is called a \emph{session}). Each session consists of two possible outcomes (called \emph{profiles}; see \cref{sec:mme} for a more detailed description of such scenarios). These scenarios are randomly generated (where \num{12} scenarios are generated to check for specific preferences, \eg male versus female, and an additional scenario is generated completely randomly). For best comparability, we leverage the \ac{MM}'s code\footnote{\url{https://www.moralmachine.net/}} to generate these scenarios. Hence, we can guarantee to have the same data distribution. We generate \num{500} sessions with \num{13} scenarios each, resulting in \num{6500} scenarios.

\subsection{Instruction Prompt Translation}
Since we want to evaluate the moral biases for the tested \acp{LLM} in a multilingual setting, we need multilingual scenarios. For this, we use the translations from the \ac{MM} as they would be presented to a user in the previously mentioned languages. We further instruct the models as described in the following paragraph. To translate this instruction, native speakers translated our initial English instruction into their native language (see \refsupplementary{\cref{tab:system-prompts} in \cref{sec:system-prompts}}). All our native speakers are fluent in English and have higher education qualifications. In an additional step, we review the translations provided to us for correctness. For this, we use a common machine translator.\footnote{https://translate.google.com/}

\subsection{Prompting}
For each \ac{LLM}, we use the model fine-tuned for chat or, if not available, an instruct model. To ensure that we can process the results automatically, we give a system prompt instructing the model to choose exactly one scenario and output only "1" or "2" as follows:
\begin{quote}
    You are a self-driving car. You will be faced with moral decisions. Each decision will be between two scenarios ("1" and "2"). You must choose one of the two scenarios. Type only "1" or "2". Do not type anything else.
\end{quote}
Importantly, we do not tell the \ac{LLM} to behave "good," find the "least fatal outcome," or "behave as one would expect from an autonomous vehicle" to not induce a confounding bias into the decisions. This is important as we want to investigate the actual bias present in the model without injecting external beliefs.

Subsequently, we use a conversational scheme as follows:
\begin{enumerate}
    \item provide a single scenario with profiles one and two prefixed by "1:" and "2:", respectively
    \item prompt the model to receive a decision ("1" or "2")
    \item repeat
\end{enumerate}
This scheme is repeated 13 times, once for each scenario, and the results are collected for further analysis. Crucially, we restart the chat session for every session of the \ac{MME}. That is, we reset the \ac{LLM}'s context such that it can only access the scenarios of the current session. If any prompt fails within a session (\eg due to a blocked prompt or an invalid response), the whole session is repeated.

Similar to the \ac{MME}, to mitigate any external biases, we randomise the scenario order and which profile is presented first or second. Moreover, we set the temperature of all models to zero (\ie configure them to select outputs deterministically) to get consistent and reproducible results.

\tabInvalidSessionsPerLanguage
\tabInvalidSessionsPerModel

\subsection{Data Analysis}
In line with the \ac{MME}, we compute the \ac{AMCE} \cite{Egami_Imai_2019} and \SI{95}{\percent} confidence interval for every factor for every language. To ensure comparability, we base our analysis on the code kindly provided by \citet{awadMoralMachineExperiment2018}. We then use these scores to investigate moral preferences and moral bias.

\paragraph{Quantitative Analysis.}
First, we analyse how well the models follow the system prompt and the prompt blockage rate (as some models have safety barriers to disallow prompts that are deemed dangerous). Furthermore, we report the \ac{MAB} and \ac{RMSE} between a model's \acp{AMCE} and the human \acp{AMCE} of the \ac{MME}. The former is the mean of the absolute values of all \acp{AMCE} measuring how pronounced the moral preferences of a model are (\ie how much its preferences deviate from taking random choices). The latter is a measure how well these moral preferences align with the \ac{MME} results.

\paragraph{Language Clustering.}
Second, we perform hierarchical clustering on the scores obtained across languages using Ward's method \cite{ward1963hierarchical} to understand how the moral biases of different languages relate. This gives us dendrogram plots similar to \citet[fig.~3a]{awadMoralMachineExperiment2018}.

\paragraph{Cultural Clustering and Biases.}
Third, we manually assign all languages to the clusters \emph{Western}, \emph{Eastern}, and \emph{Southern}, following the original classification of the \ac{MME}. To establish a connection between languages and Western, Eastern, and Southern cultures, we leverage the country clustering reported in the \ac{MME} as follows. For each language, we extract the countries where this language is an official language (using the \emph{countryinfo} Python package\footnote{\url{https://pypi.org/project/countryinfo/}}). Subsequently, we use the \ac{MME} clustering to associate those countries to cultures (Western, Eastern, and Southern) and assign each language to the culture with the most countries. For instance, Portuguese is the official language of Angola, Brazil, Portugal, and Macau, where all but the latter are deemed to be \emph{Western} countries and Macau is classified as \emph{Eastern}. Hence, Portuguese is deemed to be a \emph{Western} language. See \cref{tab:language_clusters} for all language-culture clusters and \refsupplementary{\cref{tab:language_clusters_raw} in \cref{sec:language_clusters}} for all language-country-culture associations.

Subsequently, we exploit this "manual" clustering to create preference plots similar to \citet[figs.~2 and 3b]{awadMoralMachineExperiment2018}. These radar charts allow assessing the differences in moral preferences present in any model and the cultural influence by comparing the three clusters: \emph{Western}, \emph{Eastern}, and \emph{Southern}.

%% file: 04-results.tex
\section{Results}  \label{sec:results}

\paragraph{Overall Results.}
Generally, the tested models react very differently to the experiments. \Cref{tab:invalid_sessions_per_model} shows the proportion of invalid responses per model across all languages. Due to the high non-evaluable response rate (up to \SI{100}{\percent} for all languages for Llama-2), we excluded Falcon 40B-Instruct, Falcon 180B-Chat, all Llama 2 models from all subsequent analyses. The two Falcon models did not comply with the system prompt (\eg they answered with Chinese characters instead of "1" or "2"). With Llama 2, on the other hand, the implemented safety measures blocked all of our prompts. For instance, one of the responses generated by Llama 2 was
\begin{quote}
    I cannot make decisions that result in harm or death to any living being, including pedestrians, animals, or criminals. [\dots\!]
\end{quote}
Across all models, all Llama 3 models, Gemini 1.0 Pro, and \ac{GPT} 3.5 Turbo show the lowest invalid session rate.

We also report the invalid session proportions for the individual languages across all models (\cref{tab:invalid_sessions_per_language}). We observe the highest invalid session proportions in Arabic (\SI{38}{\percent}), Korean (\SI{30}{\percent}), Portuguese (\SI{30}{\percent}) and Russian (\SI{28}{\percent}). For example, Gemini 1.0 Pro blocks many of the Arabic prompts as they are considered dangerous. We hypothesise that this finding is tied to the quality of the language-specific representation spaces. Furthermore, the models often responded with the respective characters in Korean and Chinese scenarios rather than "1" or "2". Corresponding scenarios are also labelled invalid sessions and not included in the analysis. Conversely, we record the fewest invalid sessions in English (\SI{1}{\percent}), German (\SI{4}{\percent}), and Japanese (\SI{5}{\percent}).

\paragraph{Language Clustering.}
\figDendrograms

Analogous to the methods from the \ac{MME}, we apply hierarchical clustering to the languages based on the \ac{AMCE} \cite{Egami_Imai_2019}. The corresponding results are presented in \cref{fig:dendrogram} and show how closely the individual languages are distributed concerning the inherent moral bias. Consequently, the moral bias of different languages within these clusters is similar. Due to the high invalid session rate for the models Falcon 7B-Instruct and \ac{MPT} 7B-Chat, the dendrograms from these models do not represent all languages. Since the reported hierarchical clustering from the \ac{MME} is based on the participants' countries of the experiment and ours on the languages supported by the models, a direct comparison is not possible here.

The dendrograms of Falcon 7B-Instruct (\cref{fig:dendrogram_falcon}), \ac{MPT} 7B-Chat (\cref{fig:dendrogram_mpt_7b}), and \ac{GPT} 3.5 Turbo (\cref{fig:dendrogram_gpt}), and Llama 3 8B-Instruct (\cref{fig:dendrogram_llama3_8b}) show that the \ac{AMCE} clustering of the languages is not similar to the clustering of the \ac{MME}. 
The remaining models perform better but do not entirely represent the clusters from the \ac{MME}. 

Spanish and Russian are closer to the Eastern cluster for Gemini 1.0 Pro (\cref{fig:dendrogram_gemini}). The dendrogram of Llama 3 70B-Instruct (\cref{fig:dendrogram_llama3_70b}) indicates that Spanish is closer to Portuguese and English and, therefore, part of the Western cluster. Also, Arabic is the most different from the other languages. For \ac{MPT} 30B-Chat (\cref{fig:dendrogram_mpt_30b}), English and Spanish are close to the languages of the Eastern cluster. 

Overall, none of the models replicate the clusters from the \ac{MME}. In every model, the \ac{AMCE} values of the Spanish language are closer to Western languages and not distinct enough to be seen as a different cluster.

\paragraph{Moral Bias.}
In \cref{tab:rmse_mab}, we see that \ac{MPT} 7B-Chat, 30B-Chat, and Falcon 7B-Instruct do not exhibit significant moral preferences (this does also not change when computing the \ac{MAB} for each cluster individually, see \refsupplementary{\cref{sec:additional-results}} for extended information). This indicates that these models do not inhibit a moral bias. That is, they decide randomly across all languages. We thus omit the discussion of these models due to uninteresting behaviour and refer to \refsupplementary{\cref{sec:additional-results}} for additional results. (Note that, while these models have a relatively low \ac{RMSE}, this is due to the \ac{MME} results being roughly uniformly distributed and does not indicate similar moral bias.)

\tabRmseMab

\figRadarLlamaThreeSeventyB

Opposed to that, Llama 3 70B-Instruct and Gemini 1.0 Pro exhibit pronounced moral preferences across all clusters. \Cref{fig:radar_llama3_70b_amce} reveals where these tendencies lie for Llama 3 70B-Instruct: across all cultural clusters, the model tends to spare fewer characters (\ie it prefers running over more characters). Similarly, it prefers sparing pets over humans and people with lower social status over people with higher social status. Moreover, Llama 3 70B-Instruct shows a preference towards saving passengers rather than pedestrians. In all other factors, the model does not have considerable preferences (there is a slight preference towards saving lawful people in the eastern cluster and a slight preference towards saving older people in the southern cluster). Interestingly, it does not have a preference on whether to act or not to act. This pattern dramatically diverts from human data collected in the \ac{MME}, which is also reflected in a large \ac{RMSE} of \num{0.672(3)}.

\figRadarGemini

On the other hand, Gemini 1.0 Pro has a much smaller \ac{RMSE} of \num{0.299(4)} (in fact, Gemini 1.0 Pro exhibits the smallest total \ac{RMSE} across all models). Looking at \cref{fig:radar_gemini_amce}, we can identify less pronounced moral preferences compared to Llama 3 70B-Instruct: across all clusters, Gemini 1.0 Pro prefers sparing the lawful as opposed to those crossing the street illegally and prefers saving humans over pets. For all other factors, Gemini 1.0 Pro does not exhibit significant preferences. Similarly to Llama 3 70B-Instruct, Gemini 1.0 Pro is not biased towards action or inaction.

\figRadarGpt
\figRadarLlamaThreeEightB

While these models have the largest moral bias, \ac{GPT} 3.5 Turbo and Llama 3 8B-Instruct still exhibit minor moral bias. \Cref{fig:radar_gpt_amce,fig:radar_llama3_8b_amce} show that both models prefer saving fewer characters. However, this preference is not as prevalent in the Eastern cluster for \ac{GPT} 3.5 Turbo, where the preference for saving fewer characters is reduced. Also, \ac{GPT} 3.5 Turbo shows a stark preference for saving lawful people in the Southern cluster. Both \ac{GPT} 3.5 Turbo and Llama 3 8B-Instruct have a similar \ac{RMSE} of \num{0.367(4)} and \num{0.394(4)}, respectively. As expected, these values are in between those of Llama 3 70B-Instruct, and Gemini 1.0 Pro, which are the worst and best models in terms of matching the \ac{MME}, respectively.

\paragraph{Cultural Differences.}
We also report the cultural differences regarding the moral preferences of the individual models. As before, we restrict ourselves to the models that exhibit a moral bias according to \cref{tab:rmse_mab}. The figures show the different preferences of the respective culture clusters in relation to the other clusters as z-scores across all clusters. That is, if for some models, the moral preferences are outside of the black circle, this does \emph{not} mean the model prefers sparing that attribute. Instead, the model spares this attribute \emph{more than average} over all languages. This distinction is of utter importance: a model might always prefer saving pets but is slightly more humane in the Western cluster than in the other two clusters and thus has a positive z-score in that dimension. In the following, we always refer to these \emph{relative} differences and use the phrasing "model X prefers Y in cluster Z" for brevity.

Moreover, we report the \SI{95}{\percent} confidence interval. Across all models and culture clusters, the factor \emph{gender} has the largest confidence interval, indicating that the models generally do not show cultural differences regarding saving female or male characters.

Overall, Llama 3 70B-Instruct has the smallest confidence interval across all three clusters (see \cref{fig:radar_llama3_70b}), indicating a strong cultural difference in the respective factors. Direct comparison with the results from the \ac{MME} reveals that the preferences of Llama 3 70B-Instruct differ strongly from the human preferences across all culture clusters. Furthermore, a direct comparison of the culture clusters shows that lawful actions are more likely to be saved in the Eastern cluster. In contrast, this factor is categorised in exactly the opposite way in the two remaining clusters. The Eastern cluster also clearly prefers female characters, more people, young people, higher status and pedestrians. In both the Western and Southern clusters, the other attributes are favoured. The culture clusters show apparent differences in the pedestrians versus passengers factor. While the Western cluster favours passengers, as in the \ac{MME}, pedestrians are more likely to be saved in the Eastern cluster. The Southern cluster completely resembles the average preference across all languages.

The preferences of Gemini 1.0 Pro show a large confidence interval, particularly in the factors \emph{age} and \emph{gender} (see \cref{fig:radar_gemini_zscore}). Clear cultural preferences are also visible in this model: while the Western cluster prefers no intervention, the other two culture clusters favour action by the autonomous vehicle. In the Western cluster, rescuing pets is favoured over humans, fit over unfit people and passengers over pedestrians, in contrast to the Eastern and Southern clusters. Conversely, the Eastern cluster neglects lawful behaviour, young people, and people of higher status compared to the Western and Southern clusters. Finally, both the Western and Southern clusters prefer to rescue more people. The factor \emph{gender} differs in all clusters. While the Western cluster marginally prefers to rescue male characters, the Eastern cluster favours female characters minimally. No direct preference can be determined in the Southern cluster, as the value here corresponds to the average preference.

The plots of \ac{GPT} 3.5 Turbo (\cref{fig:radar_gpt_zscore}) exhibit large confidence intervals across the clusters, especially in the factors \emph{gender}, \emph{fitness}, \emph{species}, and \emph{status}. With regard to individual preferences, all three clusters show different forms in direct comparison. In the Southern cluster, lawful behaviour is strongly preferred, whereas this is only marginally the case in the Western cluster and not at all in the Eastern cluster. In principle, tendencies in the Western cluster are not quite as strong as in the other two clusters, and, except for the factors \emph{age}, \emph{number}, and \emph{species}, they only ever lean slightly in one direction. We also note that the Eastern cluster is the only one showing greater preference towards sparing the outward properties than the other clusters. 

Llama 3 8B-Instruct also shows large confidence intervals in the factors \emph{age}, \emph{fitness}, and \emph{gender} (see \cref{fig:radar_llama3_8b_zscore}). While the Eastern cluster favours protecting pedestrians, more people, young people, and higher status, the opposite is true for the other two clusters. In this cluster, it is also preferred that the autonomously driving car performs an action, while in the Western cluster, the lane should be maintained. In the Southern cluster, on the other hand, this factor is exactly the mean value and thus expresses an indifference to this factor. In the Western cluster, female characters, animals and rule-compliant behaviour are given preference, while in the other two clusters, the opposite is the case. The Southern cluster is the only one within this model that shows a clear preference for unfit people. In both the Eastern and Western clusters, fit people are marginally preferred to be rescued. 

Compared to the \ac{MME}, cultural preferences are extremely different across all models and clusters.

%% file: 05-discussion.tex
\section{Discussion}
\label{sec:discussion}
We now discuss our findings, answer our research questions and discuss further implications. We start with a general discussion of behaviour that is consistent across all models.

First, we found that all models are slightly biased towards saving men over saving women across all clusters. However, we must note that this bias is slight.

Second, all models except Llama 8B-Instruct seem not to consider whether a character is fit or unfit or whether they are elderly or young.

Third, all models seem to prefer sparing the passengers over pedestrians, which differs from the \ac{MME} results where humans would rather spare pedestrians. One reason for this might be that humans consider the deaths \emph{their fault} and would rather sacrifice themselves for their mistakes rather than running over others. Conversely, an autonomous car is \emph{at fault} for both cases and would instead save its passengers.

Interestingly, no models in any language show a preference for action versus inaction. This is similar to the \ac{MME} results and suggests that the common issue of the trolley experiment ("If I change lanes, I am actively running over people, and thus I do nothing.") is not as prevalent as usually thought.

\subsection{Research Questions}
Given our results, we can now go back to our research questions and formulate answers for them.

\paragraph{(RQ1) Do \acp{LLM} exhibit biases reflected through their preferences when faced with moral dilemmas in  autonomous driving scenarios?}
Yes. As the various radar charts from \cref{sec:results} and \cref{tab:rmse_mab} show, the models have a moral bias to varying degrees (except for Falcon 7B-Instruct, \ac{MPT} 7B-Chat, and \ac{MPT} 30B-Chat which show no moral bias).

Llama 3 70B-Instruct is the model with the most pronounced preferences. While Gemini 1.0 Pro, \ac{GPT} 3.5 Turbo, and Llama 3 8B-Instruct have a similar moral bias, which is differently enunciated in each case, Llama 3 70B-Instruct clearly stands out regarding the reported preferences. This is unexpected regarding the other Llama models: despite the same training data, the moral bias of Llama 3 70B-Instruct and Llama 3 8B-Instruct differs significantly across the various factors. Llama 3 70B-Instruct's bias is particularly surprising since Llama 2's safety mechanisms blocks the prompts as it focuses on safety and was designed with a "no danger or harm"-policy in mind. In our analysis, however, Llama 3 is the model with the most concise moral preferences. Interestingly, Llama 3 70B-Instruct shows no utilitarian preferences across all three culture clusters and tends to run over more people rather than fewer when given the choice, as well as rather running over humans than pets. The model shows a marginal tendency to prefer deontological behaviour and save rule-following individuals in the Western and Eastern clusters.

The remaining three models (Gemini 1.0 Pro, \ac{GPT} 3.5 Turbo, and Llama 3 8B-Instruct) are similar in their moral preferences but have different degrees of preference in each case. Gemini 1.0 Pro favours deontological preferences, especially in the Southern and Western clusters. In addition, it is the only model that slightly favours humans rather than animals. This suggests that this may be an intrinsically hard-coded value. Otherwise, the model is balanced in its bias. \ac{GPT} 3.5 Turbo, on the other hand, is very indifferent across the different factors but shows a strong bias towards deontological behaviour across all clusters which is particularly pronounced in the Southern and Western cluster, similar to Gemini 1.0 Pro. In these two clusters, there is also a visible drop in the preference for the number of people rescued and here, too, the model tends to save fewer people rather than more. A similar bias, but less pronounced, is also found in Llama 3 8B-Instruct.

\paragraph{(RQ2) Is the moral bias of \acp{LLM} dependent on the prompting language?}
Yes. Both the different cultural clusters and the dendrograms clearly show that the models do not have a consistent moral bias across languages. Depending on the prompted language, the models display different response behaviours. The different dendrograms also show that all models have problems distinguishing Spanish. We hypothesise that this is probably due to the poor clustering, as maybe the main parts of the Spanish training data actually come from Spain, which follows Western values.

\paragraph{(RQ3) Does the moral bias of \acp{LLM} reflect the  culturally shaped moral dispositions of people speaking the language?}
No. Although Gemini 1.0 Pro performed best in terms of the \ac{RMSE}, its moral bias do not align with those reported in the \ac{MME}. This finding is the most surprising to us, as we previously assumed that the underlying training data represented the respective cultural moral preferences. The question now arises about whether and how language can adequately express moral preferences. A further analysis of the training data could provide a possible explanation for this behaviour.

\subsection{Comparison to \citeauthor{takemotoMoralMachineExperiment2023}'s Work}
Interestingly, we found quite different results than \citet{takemotoMoralMachineExperiment2023} for Llama 2 and \ac{GPT} 3.5 Turbo. While Llama 2 blocked all our prompts due to moral concerns, \citet{takemotoMoralMachineExperiment2023} got sufficient results. We attribute this to us using the original translations from the \ac{MME} whereas \citet{takemotoMoralMachineExperiment2023} created custom (English) descriptions. For \ac{GPT} 3.5 Turbo, they reported different moral biases than we do (comparing \citet[fig.\ 1]{takemotoMoralMachineExperiment2023} and the Western cluster of \cref{fig:radar_gpt_amce}). Most strikingly, in our setting \ac{GPT} 3.5 Turbo preferred saving less, whereas \citet{takemotoMoralMachineExperiment2023} reported a tendency towards sparing more, characters. We speculate this is due to the difference in our system prompt. While we never explicitly state that the model shall make a "right" or "morally superior" decision, \citet{takemotoMoralMachineExperiment2023} tells the model to "indicate which case is better for autonomous driving", which might bias the model. However, we have not performed further ablation studies to investigate this issue.

\subsection{Implications}
Analysing our results and answering our research questions has shown that some models have a moral bias, and when this is the case, it is not consistent across languages. As the moral bias is inconsistent across languages, interacting with \acp{LLM} can reinforce one's own culturally shaped moral biases. At the same time, this results in different response behaviours adapted to the respective languages. Resulting, the model is no longer predictable in its response behaviour. A consistent moral bias does not correspond to cultural expectations and does not represent the reality that different languages and cultures have different moral preferences. On the other hand, such a model would be more predictable and ultimately more \emph{credible} as there would be consistent responses across different languages. This could further foster human-computer interaction in a positive way as this can strengthen trust in the technology. However, the question of whether a consistent or inconsistent bias is preferable is a question of machine ethics and will not be answered here.

%% file: 07-conclusion.tex
\section{Conclusion and Future Work}  \label{sec:conclusion}

In this paper, we investigated whether \textbf{(RQ1)}~\acp{LLM} exhibit moral preferences concerning the behaviour of an autonomous car, \textbf{(RQ2)}~whether the moral bias depends on the prompted language, and \textbf{(RQ3)}~whether the moral bias reflects the respective cultural moral disposition of people speaking the language. We conclude that the answers to these questions are yes, yes, and no, respectively. Moreover, we define the term \emph{moral bias} for \acp{LLM} and define \emph{moral consistency}. We conclude that \acp{LLM} turn out not to be morally consistent in that they have different moral preferences depending on the prompted language.

While most models possess moral preferences and culture-dependent moral bias are eminent, they do not align with human biases found in the \ac{MME}. Strikingly, we found that some models, in particular Llama 3 70B-Instruct, exhibit \emph{immoral} behaviour such as running over as many characters as possible or saving pets over humans.

To summarise, we can say that one shall not entrust an \ac{LLM} with decisions that could result in harm. In particular, Llama 3 70B-Instruct shows a stark preference towards action that is against widespread ethical considerations. Moreover, one shall not expect the same moral bias of an \ac{LLM} in different languages and neither expect the moral bias of an \ac{LLM} to align with a culture's beliefs.

There are a couple of possible extension points for future work. It would be interesting to see how well an \ac{LLM} can adapt to a different culture by changing the system prompt, \eg to "You are a self-driving car in Portugal [\dots\!]." This could reveal further biases present in the model that are not revealed by language alone. Furthermore, comparing the language clusters (\cref{fig:dendrogram}) to linguistic features (\eg language families, left-to-right text, \etc) could reveal interesting patterns.

%% file: 08-limitations.tex
\section{Limitations}
As our experiments are based on the \ac{MME}, our work heavily depends on it. This results in limitations for our paper. Since, unlike in the \ac{MME}, we only look at languages and not demographic backgrounds, the \emph{Southern} cluster only consists of the Spanish language. In general, by clustering different languages into one large culture (\emph{Western}, \emph{Eastern}, and \emph{Southern}), individual subtleties of the various subordinate cultures and languages can also be lost, as with generalisation. Consequently, the clustering might be noisy. Further research should incorporate the various cultural aspects into the prompts for the \acp{LLM} at a more granular level and investigate how responses and moral bias behave. Moreover, repeating the same experiment with different system prompt formulations may reveal biases that we did not account for.

%% file: meta/acknowledgements.tex
\section*{Acknowledgements}

We would like to thank
    Abdelali Maddane,           
    Emilie Yu,                  
    Irina Rath,                 
    Sue Yoo,                    
    Yuta Noma,                  
    Zhecheng Wang,              
    Jonathan Panuelos,          
    Lily Goli,                  
    Maria Attarian,             
    Umangi Jain,                
    Umut Pajaro Velasquez, and  
    Claas Voelcker,             
for their help in translating and reviewing the prompts.

The authors gratefully acknowledge the computing time provided to them on the high-performance computer Lichtenberg at the NHR Centers NHR4CES at TU Darmstadt. This is funded by the Federal Ministry of Education and Research and the state governments participating based on the resolutions of the GWK for national high-performance computing at universities (\url{www.nhr-verein.de/unsere-partner}).

This work is partly funded under the Excellence Strategy of the Federal Government and the Länder. We thank the anonymous reviewers for their insightful comments.

%% file: 12-additional-results.tex
\section{Additional Results}  \label{sec:additional-results}

Here, we report additional results from our experiments.

\paragraph{Moral Preferences.}
We show the radar charts of the models we removed due to their low MAB scores. In the radar plots of Falcon 7B-Instruct (\cref{fig:radar_falcon_amce}), \ac{MPT} 7B-Chat (\cref{fig:radar_mpt_7b}), and \ac{MPT} 30B-Chat (\cref{fig:radar_mpt_30b_amce}) we can see that none of the models exhibits any large moral preferences. Only \ac{MPT} 30B-Chat shows a very slight preference towards sparing lawful characters in the Western cluster. Interestingly, we identify extremely large confidence intervals for Falcon 7B-Instruct (\cref{fig:radar_falcon_zscore}) which is due to Falcon 7B-Instruct having virtually no moral preferences across all cultures. Similar results can be seen for \ac{MPT} 7B-Chat and 30B-Chat (\cref{fig:radar_mpt_7b_zscore,fig:radar_mpt_30b_zscore}, respectively), although the effect is not as prevalent as for Falcon 7B-Instruct.

\figRadarFalcon
\figRadarMptSevenB
\figRadarMptThirtyB

\paragraph{Invalid Sessions, \acs{RMSE}, and \acs{MAB}.}
\Cref{tab:invalid_sessions_full} shows the proportion of invalid sessions in the datasets over the twelve different models across the ten prompted languages. Additionally, we report the RMSE and MAB of all models and the respective clusters in \cref{tab:rmse_mab_full}.

\tabInvalidSessionsFull
\tabRmseMabFull

%% file: 10-system-prompts.tex
\section{System Prompts}  \label{sec:system-prompts}

\Cref{tab:system-prompts} shows the system prompts we use to introduce the dilemma to the different \acp{LLM}. As described in \cref{sec:methods}, the instruction prompt was initially formulated in English and then translated by native speakers of the respective language. 

\tabSystemPrompts

%% file: 11-language-clusters.tex
\section{Language-Country-Culture Clusters}  \label{sec:language_clusters}

A detailed overview of the language-country-culture associations from the cultural clustering can be found in \cref{tab:language_clusters_raw}.

\tabLanguageClustersRaw